\documentclass[notoc]{JHEP3}
\usepackage{epsfig}
\def\be{\begin{equation}}  
\def\ee{\end{equation}}  
\newcommand{\ds}{{\tt DarkSUSY}}

\title{Increasing the Neutralino Relic Abundance\\ with Slepton Coannihilations: Consequences for\\
Indirect Dark Matter Detection}
\author{Stefano Profumo \\
	California Institute of Technology, Mail Code 106-38, Pasadena, CA 91125, USA\\
    E-mail: \email{profumo@caltech.edu}}
\author{Alessio Provenza \\
	SISSA/ISAS, via Beirut 2-4, 34013 Trieste, Italy and\\
          INFN, Sezione di Trieste, I-34014 Trieste, Italy\\
	E-mail: \email{provenza@sissa.it}}

\preprint{SISSA-56/2006/EP}

\abstract{We point out that if the lightest supersymmetric particle (LSP) is a Higgsino- or Wino-like neutralino, the net effect of coannihilations with sleptons is to {\em increase} the relic abundance, rather than producing the usual suppression, which takes place if the LSP is Bino-like. The reason for the enhancement lies in the effective thermally averaged cross section at freeze-out: sleptons annihilate (and co-annihilate) less efficiently than the neutralino(s)-chargino system, therefore slepton coannihilations effectively act as {\em parasite degrees of freedom} at freeze-out. Henceforth, the thermal relic abundance of LSP's corresponds to the cold Dark Matter abundance for {\em smaller} values of the LSP mass, and {\em larger} values of the neutralino pair annihilation cross section. In turn, at a given thermal neutralino relic abundance, this implies larger indirect detection rates, as a result of an increase in the fluxes of antimatter, gamma rays and neutrinos from the Sun originating from neutralino pair annihilations.}


\begin{document}

\section{Introduction}\label{sec:intro}

The fundamental nature of non-baryonic Dark Matter is one of the most pressing topics in contemporary Particle Physics. While Dark Matter is invoked in most cosmological models, and its existence, essentially inferred through gravitational effects, distribution and total abundance on cosmological scales are quantitatively established with an increasing and remarkable accuracy, little is known about what is the elementary constituent of this elusive and yet so substantial component of the cosmic budget \cite{dmreviews}. Dark Matter candidates have been proposed in several extensions of the Standard Model of particle physics, or in purely phenomenological settings, motivated, for instance, by astrophysical observations with no obvious known source counterparts; the range of particle masses proposed in the literature spans over various decades of orders of magnitude (see, {\em e.g.}, \cite{Baltz:2004tj}).

A sensible rationale to distinguish among different Dark Matter candidates emerges in the nature of the process invoked to explain why a given candidate should have a relic abundance close to the abundance of Dark Matter we infer from observations today. Along this line of reasoning, weakly interacting massive particles (WIMPs) stand as excellent prototypes. Once in thermal equilibrium with the primordial particles thermal bath in the Early Universe, WIMPs undergo a {\em freeze-out} of their number density as their pair annihilation rate becomes smaller than the Universe expansion rate. After decoupling, the WIMP number density per comoving volume remains substantially constant up to the present epoch;  their final  relic abundance crucially depends upon the WIMP pair annihilation rate. Be it a fortuitous coincidence or not, when the WIMP pair annihilation cross section is comparable to a typical weak-interactions cross section, the estimated WIMP relic abundance $\Omega_\chi h^2$ is remarkably close to the actual inferred Dark Matter abundance, $\Omega_{\rm CDM}h^2\simeq 0.110$ \cite{Spergel:2006hy} (here and in the remainder of the paper, $\Omega_i$ indicates the ratio of the mean density of species $i$ over the critical density, and $h$ the normalized value of the Hubble expansion rate today, in units of 100 km/s/Mpc).

The detailed dynamics of WIMPs freeze-out depends crucially, however, on the specific nature of the particle physics setup at hand. For instance, the relic abundance of the lightest neutralino in minimal supersymmetric extensions of the Standard Model (MSSM), a paradigmatic WIMP, varies over several orders of magnitude, questioning whether achieving $\Omega_\chi\simeq\Omega_{\rm CDM}$ is in fact ``natural'' at all. Further, the chemical decoupling of WIMPs in the Early Universe can be complicated by the concomitant, and possibly {\em coupled}, decoupling of other particle species. The occurrence of the latter scenario, known in the jargon as {\em coannihilation} \cite{coan,Griest:1990kh}, can lead to very significant effects on the final WIMP relic abundance.

The net effect of coannihilations crucially depends upon the size of the thermally averaged annihilation and coannihilation cross sections of the extra degrees of freedom participating to the freeze-out of the lightest, stable species, averaged over the total number of degrees of freedom (see the next section for a more detailed and quantitative description of coannihilations). On top of this, since the abundance of non-relativistic species in thermal equilibrium approximately follows a Maxwell-Boltzmann distribution, the importance of coannihilation processes is exponentially suppressed by the relative mass splitting between the coannihilating particles's masses and the stable particle's mass. Evidently, depending upon the particle physics setup, coannihilations can be responsible for both an {\em increase} or for a {\em decrease} in the final relic abundance of the stable species.

In the widely studied context of minimal supergravity (mSUGRA)~\cite{msugra}, a special realization of the MSSM, coannihilations are often thought to be synonym of {\em suppression} of the neutralino relic abundance. In mSUGRA the neutralino is Bino-like ({\em i.e.} the lightest mass eigenstate almost coincides with the fermionic superpartner of the hypercharge gauge boson) over most of the parameter space of the theory; Binos pair annihilate rather inefficiently in the Early Universe, as they feature a suppressed coupling to gauge bosons, and the pair annihilation into fermion-antifermion states is helicity suppressed. As noted in numerous publications, one of the few regions of the mSUGRA parameter space where $\Omega_\chi\simeq\Omega_{\rm CDM}$ is possible lies at low values of the universal scalar soft supersymmetry breaking parameter $m_0$, where the lightest neutralino is close in mass to the lightest stau \cite{elliscoan,nihei}. There, stau coannihilations enhance the effective stau-neutralino annihilation rate around neutralino freeze-out. Since staus annihilate {\em more efficiently} than Binos, the final Bino relic abundance is {\em lower} than without stau coannihilations, and can be such that $\Omega_\chi\simeq\Omega_{\rm CDM}$. Slepton coannihilations where first explicitely studied in Ref.~\cite{elliscoan}, where the (co-)annihilation cross section was approximated, in the low-velocity expansion limit, in powers of the mass-over-temperature ratio. A more accurate calculation, that takes into account the role of slepton mixing and the exact computation of the effective pair annihilation cross section, was then presented in Ref.~\cite{nihei}

Ref.~\cite{Edsjo:2003us} gave a nice example, again in the context of the mSUGRA paradigm, where slepton coannihilations {\em increase} the Bino relic abundance: if Binos resonantly annihilate through the $s$-channel resonant exchange of a heavy Higgs, adding coannihilating slepton degrees of freedom makes the total effective cross section, averaged above all degrees of freedom, smaller than without coannihilations. When Binos annihilate efficiently, slepton act as {\em parasite degrees of freedom} at the lightest supersymmetric particle (LSP) freeze-out.

Another illustrative example of the role of coannihilations in {\em enhancing} the final relic abundance of the stable particle comes from Universal Extra Dimensions (UED) \cite{Appelquist:2000nn}. In UED, the particle mass spectrum of Kaluza-Klein (KK) states is naturally highly degenerate, lying around a mass scale set by the inverse compactification scale radius $R^{-1}$. Coannihilations of the stable lightest KK particle (LKP) are therefore expected to play an important role. As pointed out in Ref.~\cite{Servant:2002aq} and in subsequent refined analyses \cite{Kong:2005hn}, for realistic spectra the effect of coannihilations is to significantly {\em increase} the relic abundance of the LKP, corresponding, in those setups, to the $B^{(1)}$ (to a good approximation the $n=1$ KK excitation of the hypercharge gauge boson). This results in a reduction of the value of $R^{-1}$ such that $\Omega_{B^{(1)}}\simeq\Omega_{\rm CDM}$ by factors as large as 2, depending upon the details of the KK spectrum. Coannihilations with KK states featuring a smaller annihilation cross section than that of the $B^{(1)}$ itself, and close in mass to it, such as right-handed KK leptons, are responsible for this effect.

In the present note we point out that, unlike the generic case of a Bino (barring fortuitous resonant annihilation channels), when the LSP is dominated by its Higgsino or Wino components, {\em i.e.} when the lightest mass eigenstate approximately corresponds to the fermionic superpartner of the neutral Higgses or of the SU(2) neutral gauge boson, the effect of slepton coannihilations is to {\em increase} the LSP relic abundance. The increase in $\Omega_\chi$ depends upon various circumstances (next-to-lightest neutralino and/or chargino coannihilations, sizable couplings to gauge bosons) that contribute to make the total effective Higgsino and Wino annihilation cross sections larger than that in presence of slepton coannihilations. As a result, the mass $m_\chi$ of neutralinos such that $\Omega_\chi\simeq\Omega_{\rm CDM}$ is pushed to smaller values, and the pair annihilation cross sections $\langle\sigma v\rangle$ to larger ones. In turn, this implies {\em larger indirect Dark Matter detection rates}, as the latter are in general proportional to the combination $\langle\sigma v\rangle/m_\chi^2$. We show that the occurrence of slepton coannihilations in scenarios where the LSP is Higgsino or Wino-like is perfectly viable, and actually takes place in several well motivated theoretical setups, where the induced degree of ``fine-tuning'' is generically not larger than that invoked in the context of the stau coannihilation region of mSUGRA.

We start our analysis with a quantitative discussion of coannihilation processes, which leads us to a guiding analytical formula. We then focus on particular phenomenological MSSM setups, motivated by several examples of GUT scale completions which can lead to similar spectra, and determine the combinations of LSP masses and mass splitting between the LSP and the coannihilating sleptons such that the thermal neutralino relic abundance saturates the Dark Matter abundance (sec.~\ref{sec:thermal}). Finally, we make use of the models featuring $\Omega_\chi\simeq\Omega_{\rm CDM}$ with slepton coannihilations to estimate the resulting enhancement in various indirect Dark Matter detection rates (sec.~\ref{sec:detect}), and summarize our conclusions (sec.~\ref{sec:conclude}).

\section{Neutralino Thermal Relic Abundance and Slepton Coannihilations}\label{sec:thermal}

The effective annihilation cross section for a system of $N$ (co-)annihilating particles $i$ of mass $m_i$ featuring a relative mass splitting, with respect to the stable lightest species $\chi$, with mass $m_\chi$, of 
\be
\Delta_i\equiv\frac{m_i-m_\chi}{m_\chi}
\ee
is given by the expression \cite{Griest:1990kh}
\be\label{eq:sigeff}
\sigma_{\rm eff}=\sum_{i,j=1}^N\ \sigma_{ij}\frac{g_ig_j}{g_{\rm eff}^2}\left(1+\Delta_i\right)^{3/2}\left(1+\Delta_j\right)^{3/2}{\rm e}^{-x(\Delta_i+\Delta_j)},
\ee
where $x\equiv m_\chi/T$, $T$ is the temperature, the $\sigma_{ij}$'s represent the various cross section of annihilation of particles $i$ and $j$ into Standard Model particles, $g_i$ stands for the number of internal degrees of freedom associated with particle $i$, and 
\be\label{eq:geff}
g_{\rm eff}\equiv\sum_{i=1}^{N}g_i\left(1+\Delta_i\right)^{3/2}{\rm e}^{-x\Delta_i}.
\ee

Eq.~(\ref{eq:sigeff}) and (\ref{eq:geff}) illustrate quantitatively the two points we alluded to in the Introduction: (1) the effective annihilation cross section relevant for the relic $\chi$ abundance can be increased or decreased as a result of extra coannihilating partners, according to the relative size of the $\chi$ pair annihilation cross section and the (co-)annihilation cross section of the coannihilating partners; (2) the effect of coannihilations depends exponentially upon the ratio $\Delta_i$, times a factor accounting for the actual $\chi$ freeze-out temperature.

In the present context, we deal with a situation where the effective $\chi$ pair annihilation cross section is larger than that of its coannihilating partners, the sleptons. The Wino and Higgsino effective annihilation cross section without slepton coannihilations actually results from a combination of the various contributing (co-)annihilation cross sections of the lightest neutralino and the lightest chargino, as well as, for the case of a Higgsino LSP, with the next-to-lightest neutralino (this depends on the neutralino $\chi_i$ and chargino $\chi^\pm_i$ mass spectrum: in the case of a Wino LSP, $m_{\chi_1}\simeq m_{\chi^\pm_1}\simeq M_2$, and in the case of a Higgsino LSP  $m_{\chi_1}\simeq m_{\chi_2}\simeq m_{\chi^\pm_1}\simeq \mu$). In most MSSM realizations, the resulting overall Wino and Higgsino effective cross section is {\em larger} than the slepton pair annihilation cross sections and of the slepton-neutralino and slepton-chargino coannihilation cross sections.

In this context, it is easy to draw a rough theoretical estimate of the relative enhancement of the thermal relic abundance $\Omega_\chi$ in presence of parasite degrees of freedom associated to a set of coannihilating particles $\tilde L$ (in our case, the sleptons), assumed to lie all at the same mass scale $m_{\tilde L}$, for simplicity. We shall hereafter indicate the relative mass splitting $\Delta_{\tilde L}\equiv(m_{\tilde L}-m_\chi)/m_\chi$. Suppose the total effective neutralino annihilation cross section (including, in the case of Higgsinos and of Winos, the contribution of the next-to-lightest neutralino and/or of the lightest chargino) without the contribution of parasite particles $\tilde L$ ($\Delta_{\tilde L}\gg1$) is given by $\sigma^0_{\rm eff}=\sigma_{\chi\chi}$. The assumption that the extra coannihilating degrees of freedom associated to $\tilde L$ act as ``{\em parasite}'' degrees of freedom quantitatively amounts to have $\sigma_{\chi\chi}\gg\sigma_{\chi\tilde L},\sigma_{\tilde L\tilde L}$, where we indicate with $\sigma_{\chi\tilde L},\sigma_{\tilde L\tilde L}$ the $\tilde L$ coannihilation and self-annihilation effective cross sections, respectively. Denoting with $g^0_{\rm eff}$ the effective degrees of freedom when the $\tilde L$ particles are much heavier than the LSP ($\Delta_{\tilde L}\gg1$), the new effective total annihilation cross section $\sigma_{\rm eff}$ can be expressed as a function of the effective degrees of freedom $g_{\rm eff}$ including the $\tilde L$ particles as
\be
\sigma_{\rm eff}\simeq\sigma_{\rm eff}^0\left(\frac{g^0_{\rm eff}(x_{\rm f.o.})}{g_{\rm eff}(x_{\rm f.o.})}\right)^2
\ee
where $x_{\rm f.o.}$ corresponds to temperatures around the $\chi$ freeze-out, $T_{\rm f.o.}\approx m_\chi/25$. Conversely, the relative enhancement in the $\chi$ relic abundance will be approximately given by
\be
\frac{\Omega_\chi}{\Omega^0_\chi}\simeq\left(\frac{g_{\rm eff}(x_{\rm f.o.})}{g^0_{\rm eff}(x_{\rm f.o.})}\right)^2\approx\left(\frac{g^0_{\rm eff}(x_{\rm f.o.})+ g_{\tilde L} \left(1+\Delta_{\tilde L}\right)^{3/2}{\rm e}^{-x_{\rm f.o.}\Delta_{\tilde L}})}{g^0_{\rm eff}(x_{\rm f.o.})}\right)^2,\label{eq:master}
\ee
where we indicate with $g_{{\tilde L}.}$ the total number of internal degrees of freedom associated with the $\tilde L$ particles. In the case under investigation here, $g^0_{\rm eff}(x_{\rm f.o.})\approx6,8$ in the Wino and Higgsino case, respectively (recalling that every neutralino carries 2 internal degrees of freedom, while every chargino carries 4, and neglecting the mass splitting between the lightest neutralino and next-to-lightest neutralino and the lightest chargino), while $g_{{\tilde L}}=2,4,18$ when only the SU(2) singlet third generation slepton, the  SU(2) doublet third generation sleptons and all the sleptons are coannihilating (for conciseness, we shall indicate in the figure labels throughout the present paper the quantity $g_{{\tilde L}}$ simply with $g$). 

To make quantitative estimates of the neutralino relic density enhancement, we need a specific MSSM setup; we then compute $\Omega_\chi h^2$ numerically, making use of publicly available codes (namely, \ds \cite{Gondolo:2004sc} and {\tt micrOMEGAs} \cite{Belanger:2006is}). To this extent, we consider, for the Higgsino-like neutralino case, a value of $\mu=800$ GeV, $M_1=5\mu$ and an mSUGRA-motivated hierarchy among the gaugino masses at the low-energy scale ($M_2=2M_1$, $M_3=6M_1$). For the Wino-like case, we resort instead to a minimal anomaly-mediated supersymmetry breaking (mAMSB) inspired setting \cite{amsb} for the gaugino masses ($M_1=3M_2$, $M_3=8M_2$), and set $M_2=1.2$ TeV and $\mu=5M_2$. In both cases we set $\tan\beta=20$ and $m_A,\ m_{\rm Squarks}\approx 10\times m_{\chi}$. The last choice is motivated by avoiding, in the following discussion, spurious effects deriving from squark coannihilations. Unlike sleptons, strongly interacting squarks potentially feature a larger effective cross section than the neutralino/chargino systems of Wino and Higgsino-like neutralinos; therefore the effect we discuss here does not apply when the LSP coannihilates with squarks. Moreover, we take a large value for $m_A$ in order to forbid resonant neutralino annihilations through $s$-channel $A$ exchange diagrams, which can potentially blur the effect under investigation in the present analysis. The mass of sleptons which are not assumed to coannihilate is also set to the same value as that of $m_{\rm Squarks}$.

The setup we refer to is motivated by several theoretical studies discussed in the literature. In the context of mSUGRA, the LSP can be Higgsino-like in the Focus Point/Hy\-per\-bo\-lic Branch region, at very large values of the common supersymmetry braking scalar mass $m_0$ \cite{Baer:2005ky}. In this case, however, sleptons are very heavy and cannot coannihilate with the LSP. Going beyond mSUGRA, and relaxing some of the universality assumptions on the soft supersymmetry breaking terms drastically changes the situation. 

Ref.~\cite{Baer:2005bu} addressed the case of non-universality in the soft breaking Higgs masses (non-universal Higgs mass, NUHM, model); in that context, a Higgsino-like LSP is naturally achieved for arbitrarily small sfermion masses, and slepton coannihilations with Higgsinos can very well take place. As pointed out in \cite{Baer:2005bu}, in NUHM models the usual mSUGRA hierarchy $m_{\tilde \tau_R}<m_{\tilde \tau_L}$ between right and left handed sfermions can be subverted, and left-handed sleptons can be lighter than their right-handed counterparts, see {\em e.g.} their Fig.~11. Special values of the Higgs soft breaking masses even allow for a quasi-degeneracy of the full slepton spectrum. In this respect, one can therefore expect several slepton coannihilation scenarios: the right-handed stau alone ($g\approx2$), all right-handed sleptons ($g\approx6$), left-handed third generation sleptons ($g\approx 4$), all left-handed sleptons ($g\approx12$) or even the extreme situation of all sleptons ($g\approx18$).

Relaxing the universality of gaugino masses at the grand unification (GUT) scale, again within mSUGRA, also naturally leads to a Higgsino LSP \cite{nugmh}, as well as to a Wino-like LSP \cite{nugmw} (non-universal gaugino mass (NUGM) models; see also \cite{othernugm}). Values of $\mu$ smaller than $M_{1,2}$ can be achieved setting $M_3$ smaller than $M_1=M_2=M_{1/2}$ at the GUT scale (where $M_{1/2}$ stands for the mSUGRA universal gaugino soft supersymmetry breaking mass parameter), through renormalization group evolution \cite{nugmh}. Retaining the universality assumption in the scalar sector, in the NUGM model the lightest slepton is the right handed stau, with the lightest (right-handed) smuon and selectron relatively close in mass. One therefore expects a value of $g$ between 2 and 6, for full slepton coannihilations.

\FIGURE[!t]{
\mbox{\hspace*{-0.5cm}\epsfig{file=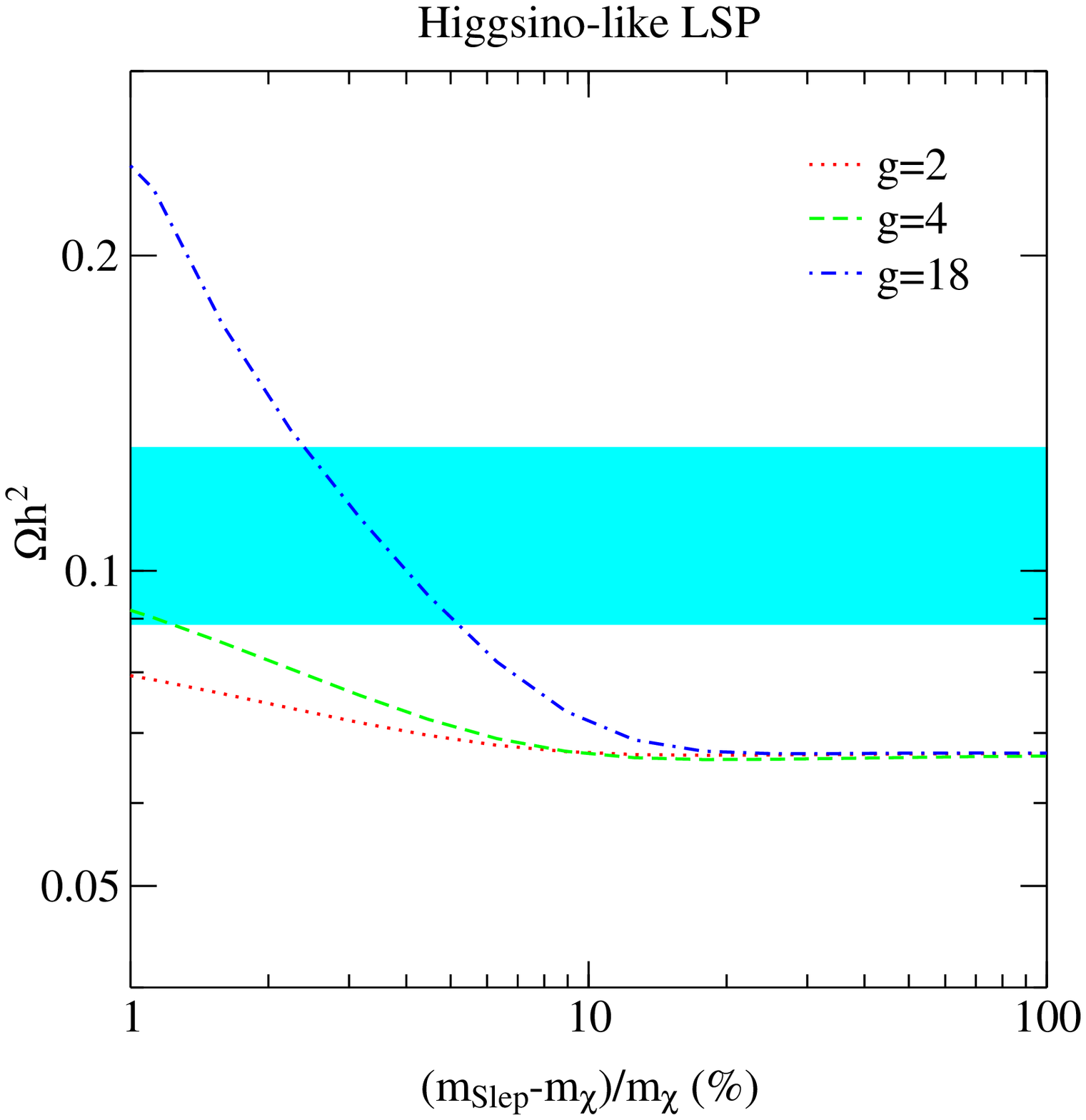,width=7.5cm}\qquad
\epsfig{file=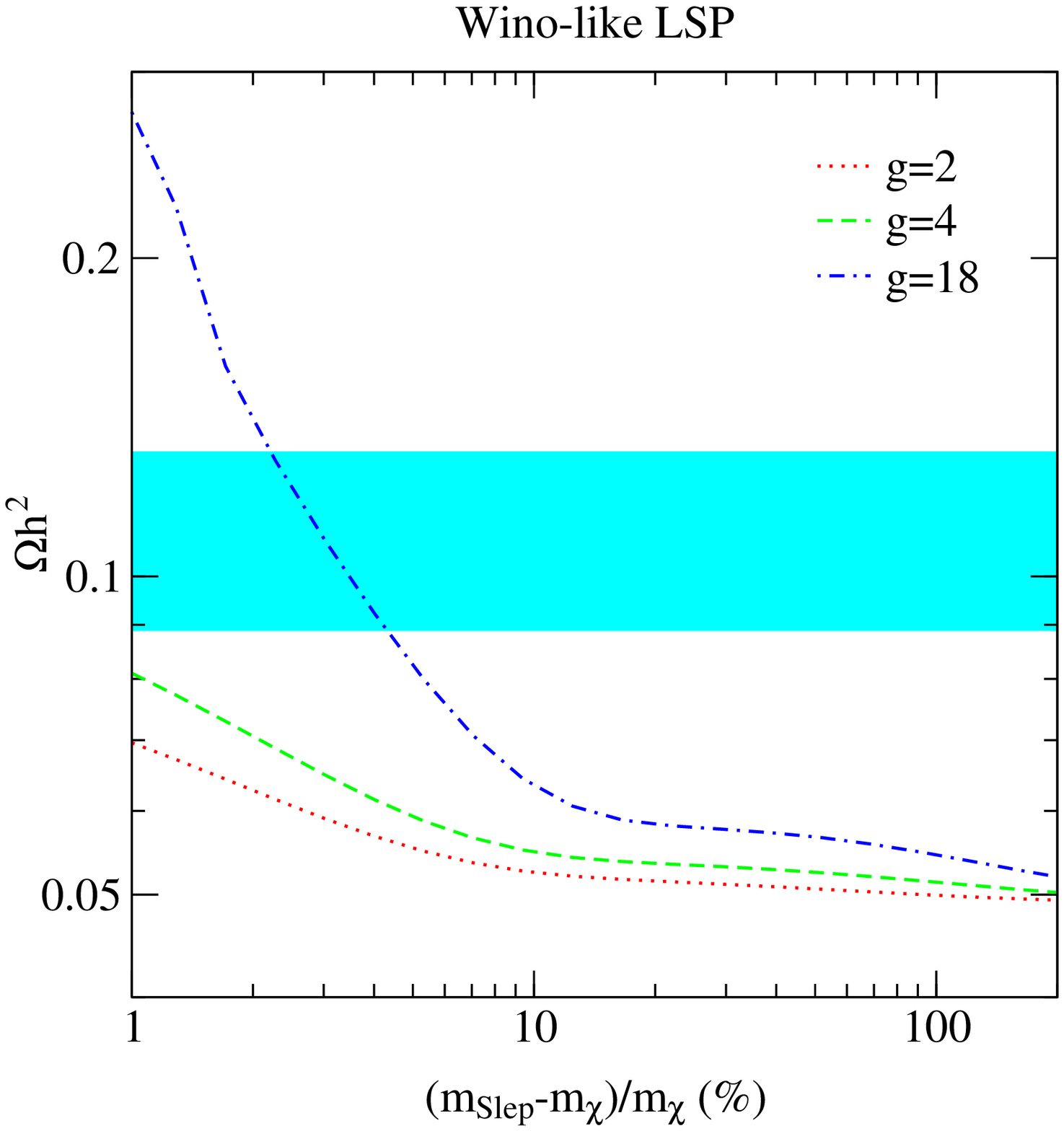,width=7.5cm}}
\caption{\label{fig1}The  LSP thermal relic abundance, as a function of the percent mass splitting between the LSP and the coannihilating sleptons. The case of a $m_\chi=800$ GeV Higgsino-like neutralino is shown in the left panel, while that of a $m_\chi=1200$ GeV Wino-like neutralino is featured in the right panel. The horizontal bands indicate the range of $\Omega_\chi h^2$ corresponding to the abundance of cold Dark Matter inferred by the WMAP team for a $\Lambda$CDM cosmology at 2-$\sigma$ \cite{Spergel:2006hy}. The label $g$ stands for the number of coannihilating slepton degrees of freedom (see the text for more details).}}
Our reference setup for Wino-like neutralinos will however be that of mAMSB \cite{amsb}. The nature of the LSP in mAMSB scenarios is determined by the gaugino soft supersymmetry breaking masses being proportional to the associated gauge group beta functions times the gravitino mass, and features, typically, a Wino-like LSP (a Higgsino-like LSP is also possible, through the analogous of the focus point effect in mSUGRA). The problem of negative slepton mass squared is solved, in the context of mAMSB, through a common phenomenological scalar mass parameter $m_0^2$. Within this setup, the lightest sfermion is the right-handed stau, although in some parameter space regions the two staus can be significantly close in mass. Selectrons and smuons always tend to be very close in mass. The absolute value of $m_0^2$ allows one to naturally obtain slepton coannihilations with a Wino-like LSP. Relaxing, here, the assumption of universality for the phenomenological parameter  $m_0^2$ easily entails all possible slepton coannihilation patterns, suitably adjusting {\em e.g.} the value of the left and right handed parameters $(m_0^2)_{L,R}$ for the slepton sector.

We thus conclude that slepton coannihilations with Higgsinos and Winos are a perfectly viable possibility in several theoretically motivated supersymmetric setups. For computational ease, we resort here to a handier low-energy scale parameterization, which, however, captures the main features of the general problem in more generic scenarios.

In Fig.~\ref{fig1} we show the neutralino relic density, computed with the {\tt micrOMEGAs} code, in the case of a Wino ($g^0_{\rm eff}(x_{\rm f.o.})=6$) and in the case of Higgsino ($g^0_{\rm eff}(x_{\rm f.o.})=8$), as a function of $\Delta_{\tilde L}$, for the two Higgsino- and Wino-like neutralino reference supersymmetric setups discussed above. In both cases the injection of the parasite slepton degrees of freedom (we focus, here and in what follows, on the cases $g=2,4$ and 18) enhances the thermal relic density up to values in the 2-$\sigma$ WMAP allowed region. The increase in the relic abundance, down to a relative mass splitting of the order of 1\%, can be as large as a factor 5, when all sleptons participate in the coannihilation process. 

\FIGURE[!t]{
\mbox{\hspace*{-0.5cm}\epsfig{file=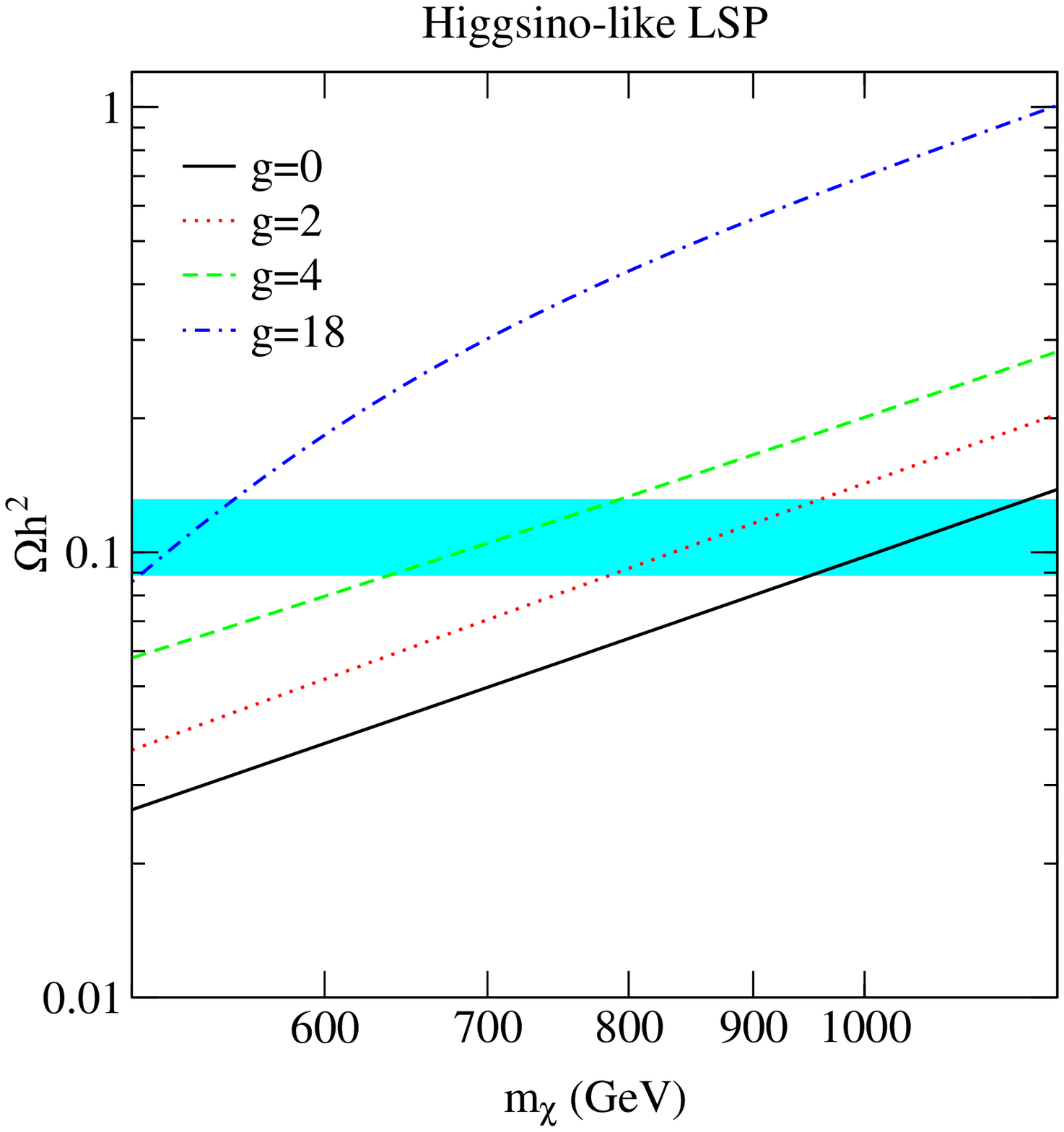,width=7.5cm}\qquad
\epsfig{file=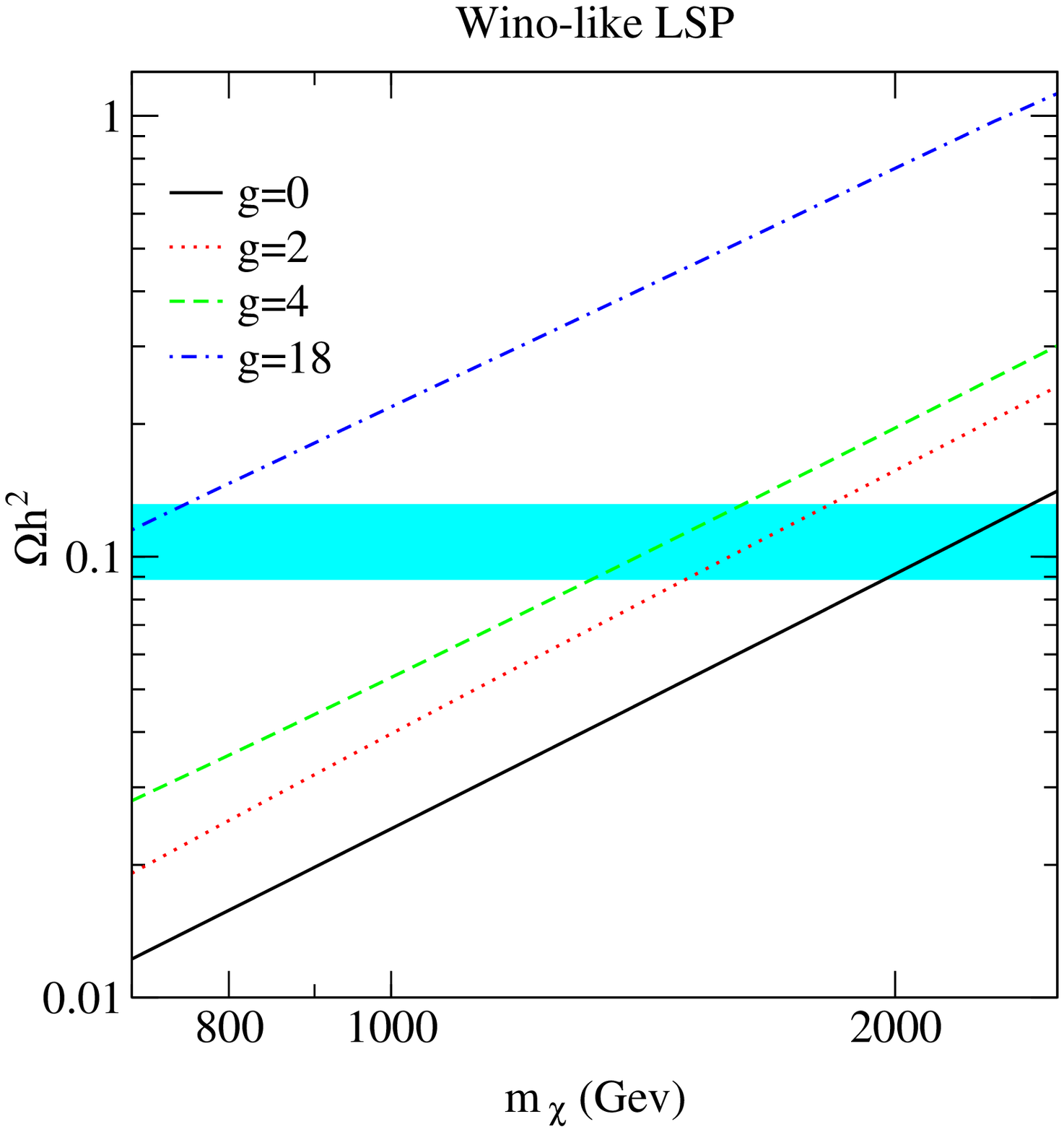,width=7.5cm}}
\caption{\label{fig2}
The thermal relic abundance $\Omega_\chi h^2$ as a function of the lightest neutralino mass in the extreme case of vanishing mass splitting between the LSP and the (coannihilating) sleptons, for a Higgsino-like neutralino (left panel) and a Wino-like neutralino (right panel). The horizontal band indicates the range of $\Omega_\chi h^2$ corresponding to the abundance of cold Dark Matter inferred by the WMAP team \cite{Spergel:2006hy}.}}

Although qualitatively the approximate formula given in Eq.~(\ref{eq:master}) reproduces the correct trend found in the numerical results shown in Fig.~\ref{fig1}, for some values of the relative mass splitting we do find quantitative differences. The latter originate from the assumptions used to derive the analytical approximation. In particular, in the theoretical estimate of Eq.(\ref{eq:master}) we  neglect the annihilation  and coannihilation cross sections for sleptons, hence overestimating the thermal relic density enhancement; secondly, we set $g^0_{\rm eff}=6$ (or $8$)  while the  real value is generically smaller; lastly, assuming a putative value $x_{\rm f.o.}=25$ does not always matches the actual numerical value for the freeze-out temperature. In any case, we stress that Eq.~(\ref{eq:master}) provides us with a useful analytical insight and a qualitative prediction for the effect we are focusing on in the present analysis.

We also point out that in the right panel of Fig.~\ref{fig1}, {\em i.e.} for the Wino case, we find a a non negligible enhancement also for values of the mass splitting well beyond the level of $\approx10$\%, where one would expect some effect related to coannihilations from the discussion above and from Eq.~(\ref{eq:master}). This fact is traced back to the kinematic enhancement in the chargino and neutralino pair annihilation cross section in presence of lighter sleptons, and has actually nothing to do with slepton coannihilations.
 
\FIGURE[!t]{
\mbox{\hspace*{-0.5cm}\epsfig{file=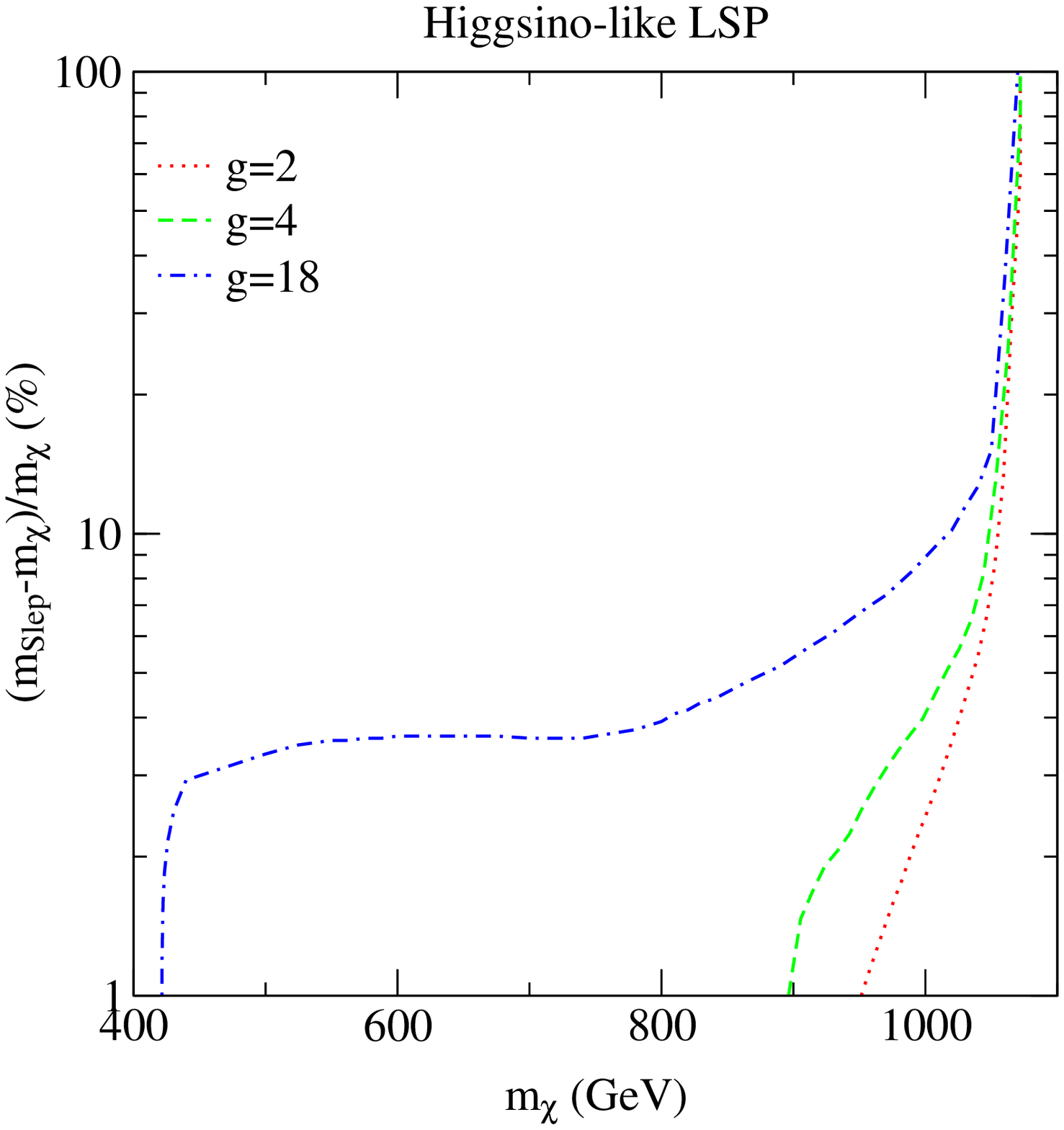,width=7.5cm}\qquad
\epsfig{file=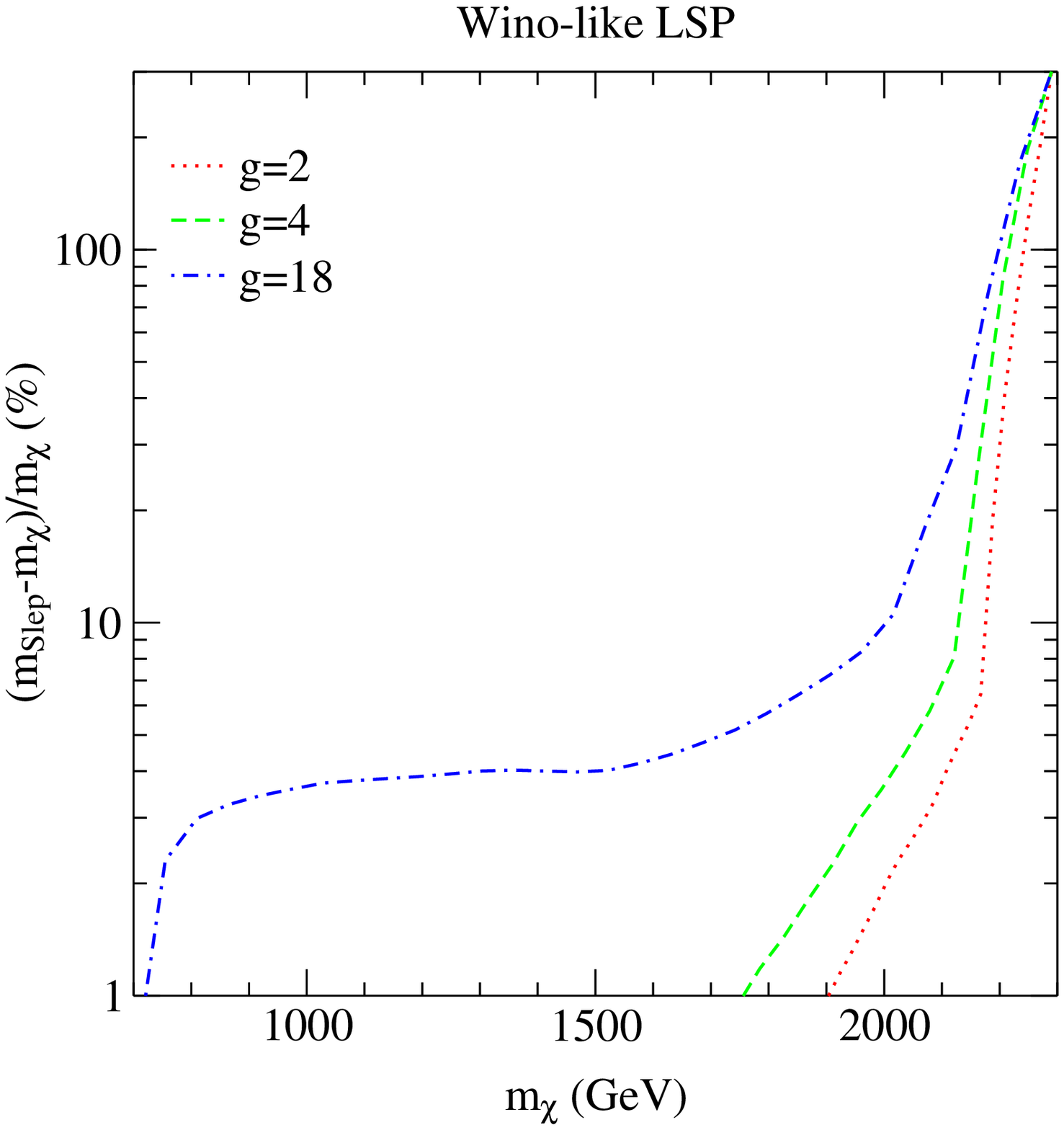,width=7.5cm}}
\caption{\label{fig3} Isolevel curves of the lightest neutralino relic
abundance corresponding to a neutralino thermal relic abundance $\Omega_\chi h^2$ equal to the central value for $\Omega_{\rm CDM}h^2\simeq 0.110$,
 in the plane defined by the LSP mass versus the relative mass splitting between the LSP and the (coannihilating) sleptons, for a Higgsino-like neutralino (left panel) and for a Wino-like neutralino (right panel).}}
In Fig.~\ref{fig2} we show the neutralino thermal relic abundance as a function of the LSP mass in the extreme case of coannihilating particles completely degenerate, in mass, with the lightest neutralino. Hereafter, the neutralino mass is varied keeping the ratios between $\mu$ and the gaugino soft supersymmetry breaking masses fixed at the values corresponding to our two reference models. The other supersymmetric parameters are kept fixed. In this way, spurious effects originating from the details of the neutralino composition are expected to be minimized, while our hypothesis on the supersymmetric setup are kept simple enough. 

We notice that the upper bound on the LSP mass from its thermal relic abundance is significantly lowered. For example even in the case of a Higgsino-like neutralino coannihilating with the third generation right handed slepton alone, the upper limit on the LSP mass is  about 20\% smaller with respect to the case whitout coannihilation; in the most extreme case of coannihilations with all sleptons, the effect amounts to a suppression in the upper limit on the LSP mass of a factor close to 4.
 
We clarify and detail on this point in Fig.~\ref{fig3}, where we plot, in the $(m_\chi,\,\Delta_{\tilde L})$ plane, the isolevel curves at $\Omega_{\rm CDM} h^2\simeq\Omega_\chi h^2\simeq0.110$. Points on the curves shown feature the ``right'' neutralino thermal relic abundance. In the case of mass splitting of 1\% and all sleptons coannihilating, the upper bound on the LSP mass is about one half of the LSP mass value without sleptons coannihilations, both for the Higgsino- and Wino-like case;  this effect is less spectacular, but also appreciable, in the case with  third generation right handed slepton coannihilations ($g=2$) or with third generation left handed sleptons coannihilations ($g=4$).

\section{The Enhancement of Indirect Dark Matter Detection Rates}\label{sec:detect}

Numerous theoretical and experimental efforts have been directed in recent years to the possibility of inferring the existence of galactic (or even extra-galactic) particle Dark Matter through the presence of exotic ``signatures'' in stable end-products of Dark Matter pair annihilations (for reviews on the topic see {\em e.g.} Ref.~\cite{dmreviews}). In particular, Dark Matter pair annihilations in the Galactic Halo can yield sizable positron and antiproton fluxes, which might be disentangled from the cosmic ray secondary and tertiary backgrounds (see {\em e.g.} \cite{Hooper:2004bq, Profumo:2004ty}); low-energy antideuterons are also among the stable hadronization products of pair annihilations of neutralinos, or other WIMPs, in the Halo, and suffer from a relatively small background \cite{Baer:2005tw}. Neutralinos captured in the core of the Sun or of the Earth through scattering with ordinary matter and subsequent gravitational collapse, can pair annihilate and produce a coherent and possibly detectable flux of energetic neutrinos \cite{dmreviews}. Finally, gamma rays from the decay of hadrons produced in pair-annihilation of neutralinos, or promptly produced in loop-suppressed processes at a monochromatic energy, are also among promising indirect detection methods \cite{dmreviews}.

\FIGURE[!t]{
\mbox{\hspace*{-0.5cm}\epsfig{file=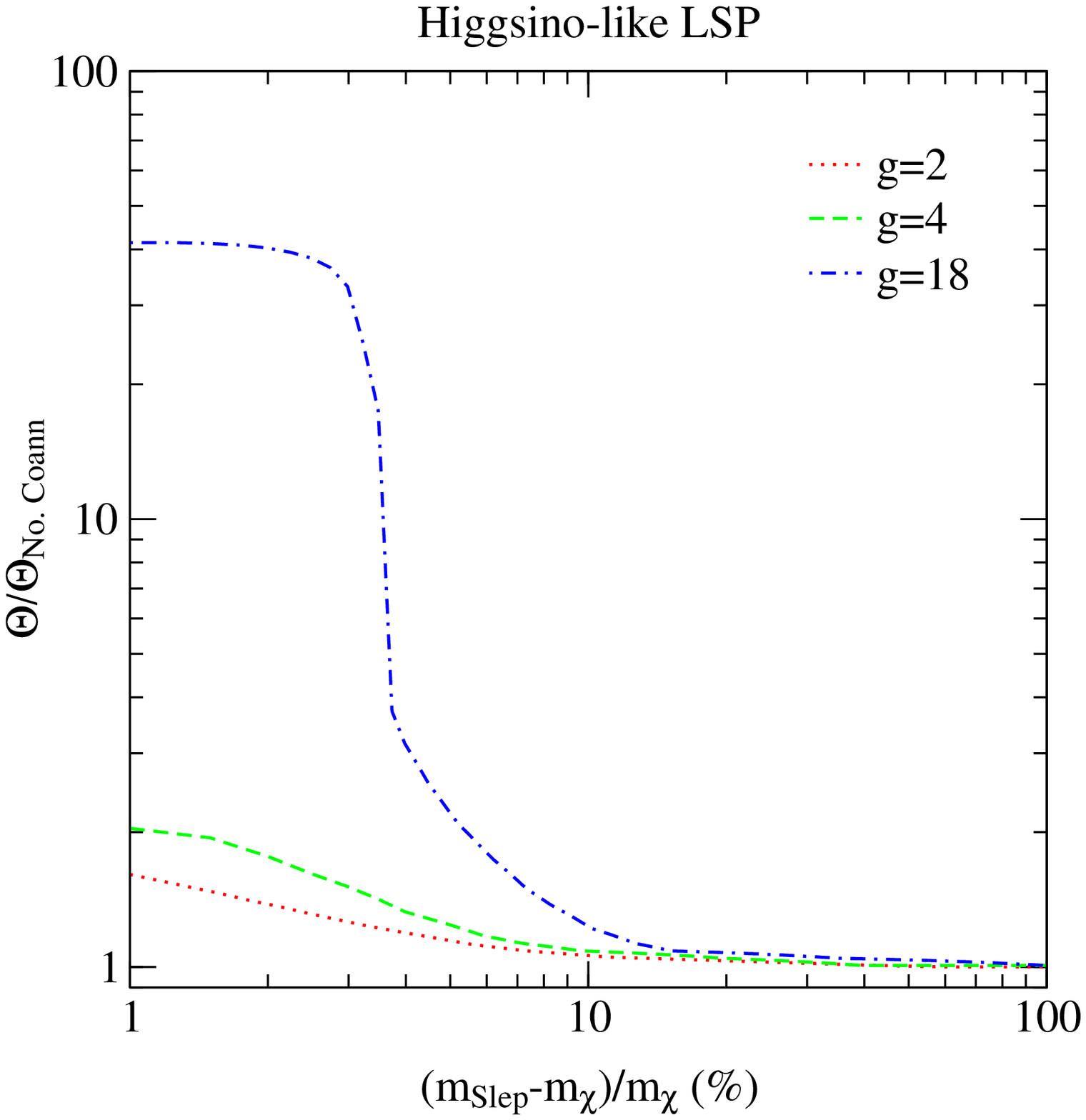,width=7.5cm}\qquad
\epsfig{file=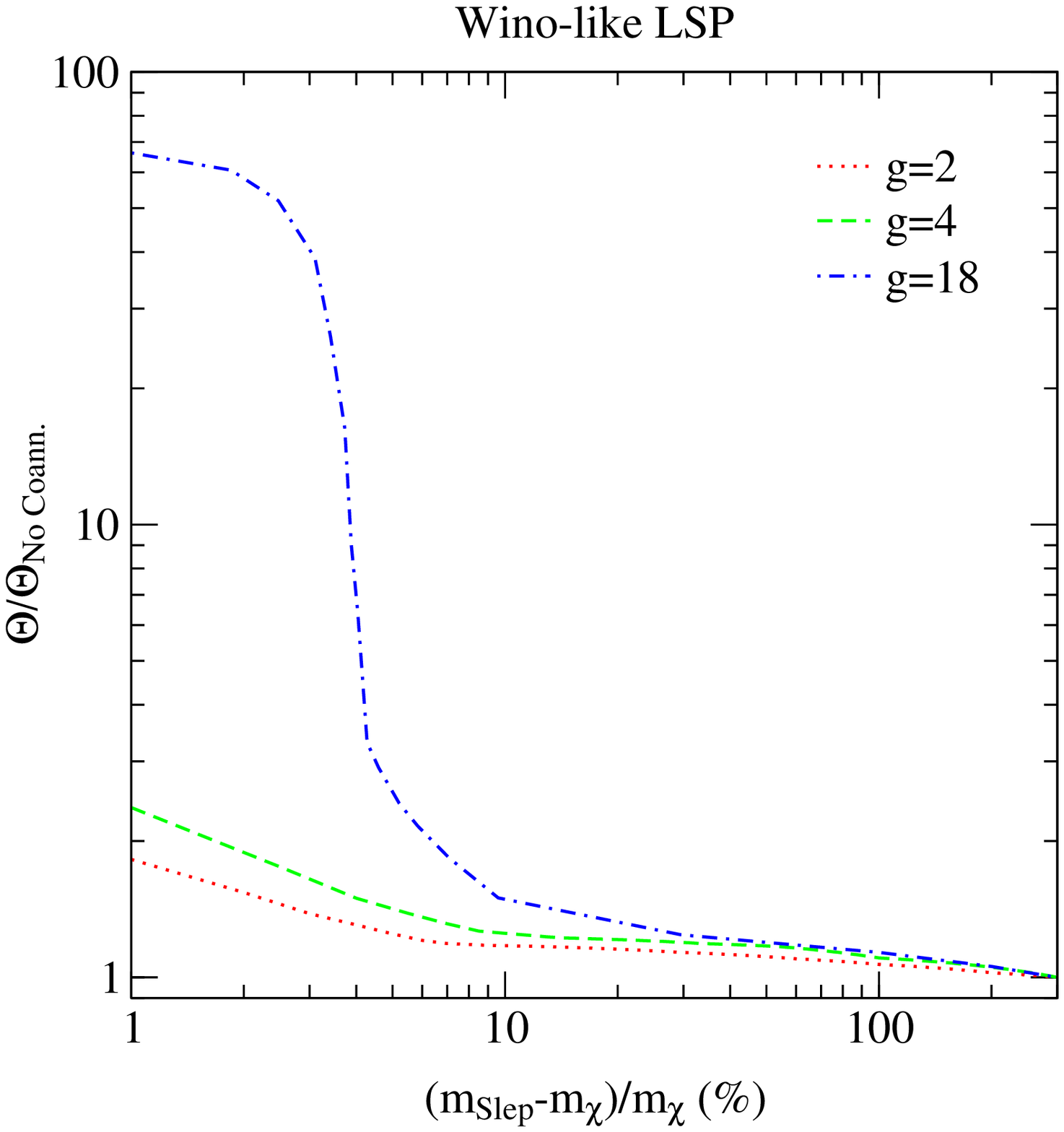,width=7.5cm}}
\caption{\label{fig4}
The relative enhancement (with respect to the asymptotic value with decoupled heavy sleptons) in the quantity $\Theta\equiv\langle\sigma v\rangle/m_\chi^2$, relevant for all indirect Dark Matter detection rates, as a function of the slepton-LSP percent mass splitting; for  the case of a Higgsino-like neutralino ({left panel}) and  a Wino-like neutralino ({right panel}). The models displayed are those at $\Omega_\chi h^2=0.110$ singled out in Fig.~\protect{\ref{fig3}}, with the same
sample choice of parameters and the same line-type and color coding.}}
The most crucial particle physics quantity involved in the assessment of generic indirect particle Dark Matter detection rates is the pair-annihilation rate today ({\em i.e.} at ``zero temperature'') times integrals involving the number density of Dark Matter pairs. In turn, this latter quantity, for a fixed Dark Matter {\em density} profile, scales with the inverse square of the Dark Matter particle mass. In Fig.~\ref{fig4} we show the enhancement of the quantity $\Theta=\left<\sigma v\right>/m^2_\chi$, computed with \ds~\cite{Gondolo:2004sc},  with respect to the case without slepton coannihilations, as a function of the relative percent mass splitting between the LSP mass and the coannihilating particle masses. We show, again, the models featuring a neutralino thermal relic abundance $\Omega_\chi h^2\simeq0.110$ determined in the previous Fig.~\ref{fig3}.

We wish to emphasize that in the extreme case of all sleptons coannihilating, the generic enhancement with respect to the asymptotic values is remarkable. We hence expect indeed significant improvements in the prospects for Dark Matter indirect detection within the present setup. In what follows, we briefly review the actual detailed size of the enhancement expected for several different indirect detection techniques.

\FIGURE[!t]{
\mbox{\hspace*{-0.5cm}\epsfig{file=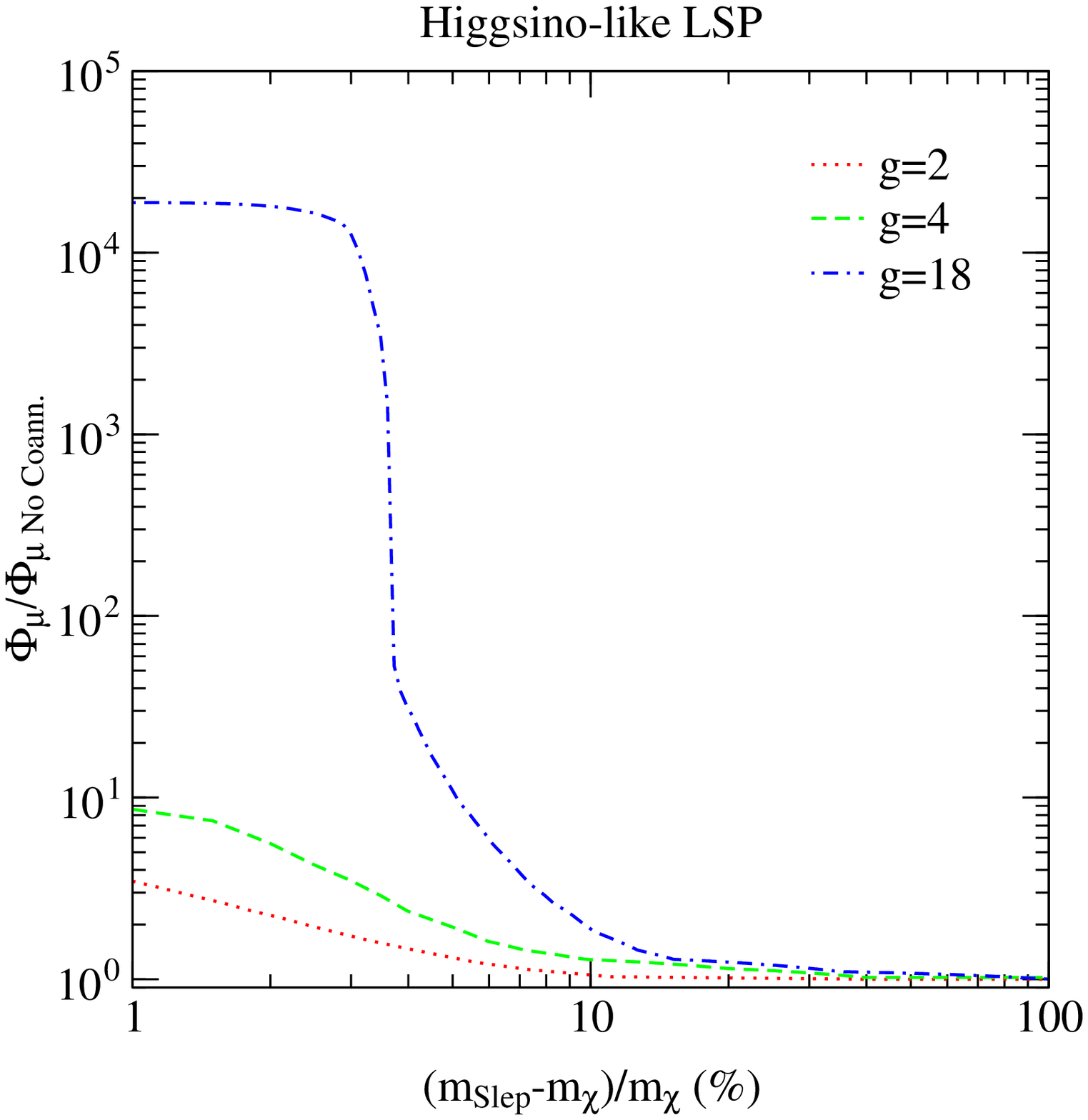,width=7.5cm}\qquad
\epsfig{file=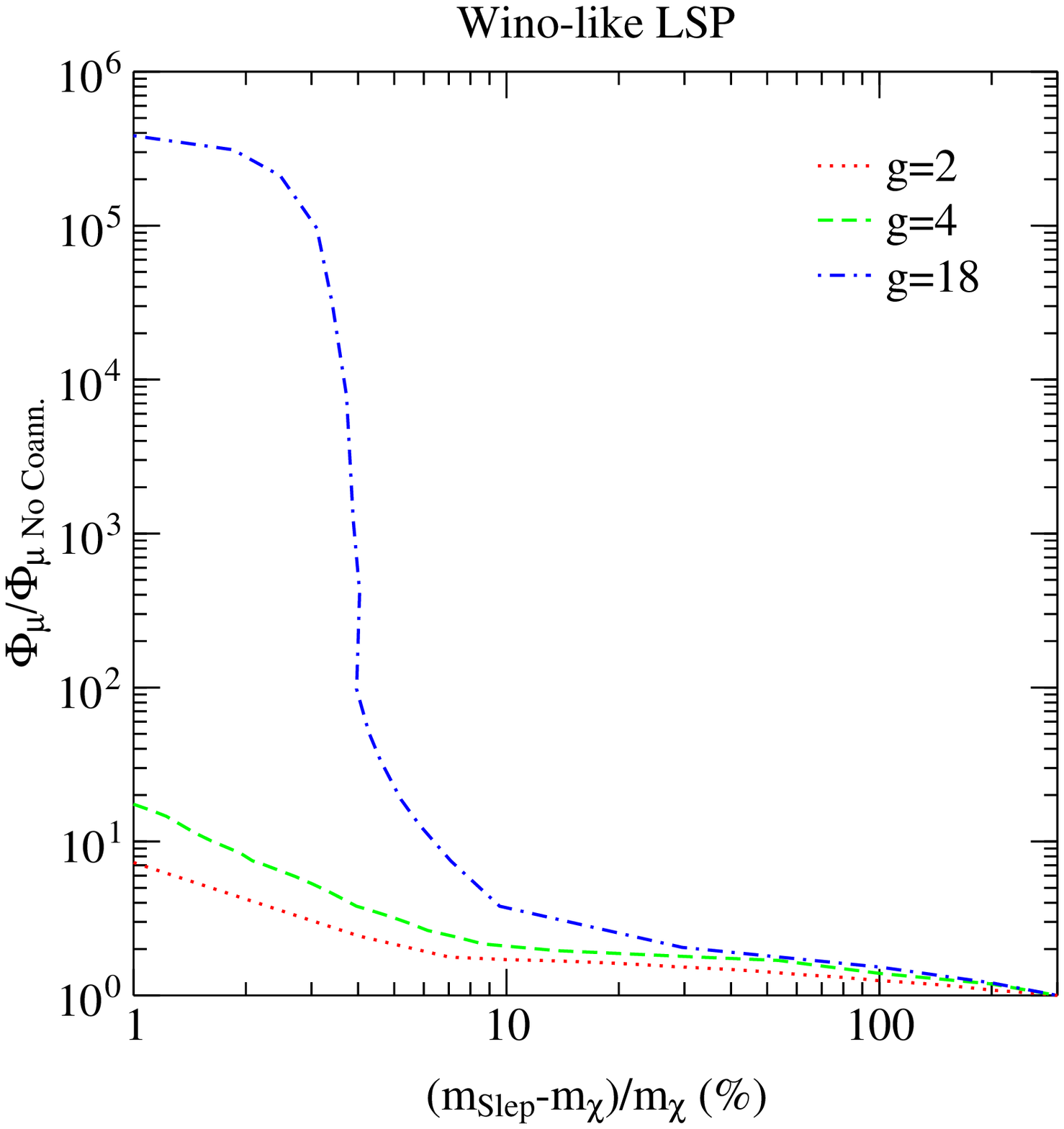,width=7.5cm}}
\caption{\label{fig5}
The relative enhancement (with respect to the asymptotic value with decoupled heavy sleptons) in the muon flux induced by  energetic neutrinos from the Sun ($E_\mu>50$ GeV),  as a function of the slepton-LSP percent mass splitting, for  the case of a Higgsino-like neutralino ({left panel}) and  a Wino-like neutralino ({right panel}). The models displayed are those at $\Omega_\chi h^2=0.110$ singled out in Fig.~\protect{\ref{fig3}}, with the same sample choice of parameters and the same line-type and color coding.}}
In Fig.~\ref{fig5} we show the enhancement of the muon flux $\Phi_{\mu}$ induced by neutralinos annihilating in the core of the Sun and producing a flux of energetic neutrinos with respect to the case without slepton coannihilations, again as a function of the relative percent mass splitting. We employ a relatively large muon energy threshold, namely 50 GeV, appropriate for ${\rm km}^3$ neutrino telescopes such as IceCube~\cite{Achterberg:2005fs}. The enhancement in the signal, showed in Fig.~\ref{fig5},  is larger than the enhancement in the annihilation cross section. To understand this effect we recall that the magnitude of the neutrino flux depends upon two quantities: the Sun capture rate, mostly driven by the spin-dependent LSP-nucleons scattering cross section, and the flux of neutrinos produced per neutralino annihilation; the total enhancement accounts for both these factors. When the neutralino mass is reduced, the role of the off-diagonal entries related to electro-weak symmetry breaking effects in the neutralino mass matrix and in the LSP composition becomes more and more important (intuitively, the relevance of the mixing induced by the mentioned entries roughly scales as $(m_W/m_\chi)^2$). As a result, a larger gaugino-higgsino mixing is expected at smaller neutralino mass. In particular, this results in a net increase in the quantity $|N_{13}|^2-|N_{14}|^2$, which enters in the $\chi\chi Z^0$ vertex, and drives an enhancement, at small neutralino masses, of more than one order of magnitude in the neutralino spin-dependent cross section off nucleons. The gaugino fraction, in the Higgsino case, and the Higgsino fraction, in the Wino case, are however always greatly suppressed, typically lying around $10^{-4}$.

In the particular models we consider here, both the asymptotic and the fully enhanced values for the muon flux do not give a signal which might be detectable with IceCube~\cite{Achterberg:2005fs}; this mostly depends upon the size of the product of the Higgsino and gaugino fractions of the lightest neutralino: to avoid spurious effects ({\em e.g.} a Bino component, and the consequent associated neutralino degrees of freedom, in the Higgsino-like case) we picked models with a suppressed spin-dependent coupling to matter. However, we explicitly checked that allowing for a larger Higgsino-gaugino mixing, the enhancement in the flux of muons in neutrino telescopes caused by the occurrence of slepton degrees of freedom at neutralino freeze-out can indeed be crucial, and make models otherwise giving a hopelessly small neutrino flux from the Sun, detectable with IceCube.

\FIGURE[!t]{
\mbox{\hspace*{-0.5cm}\epsfig{file=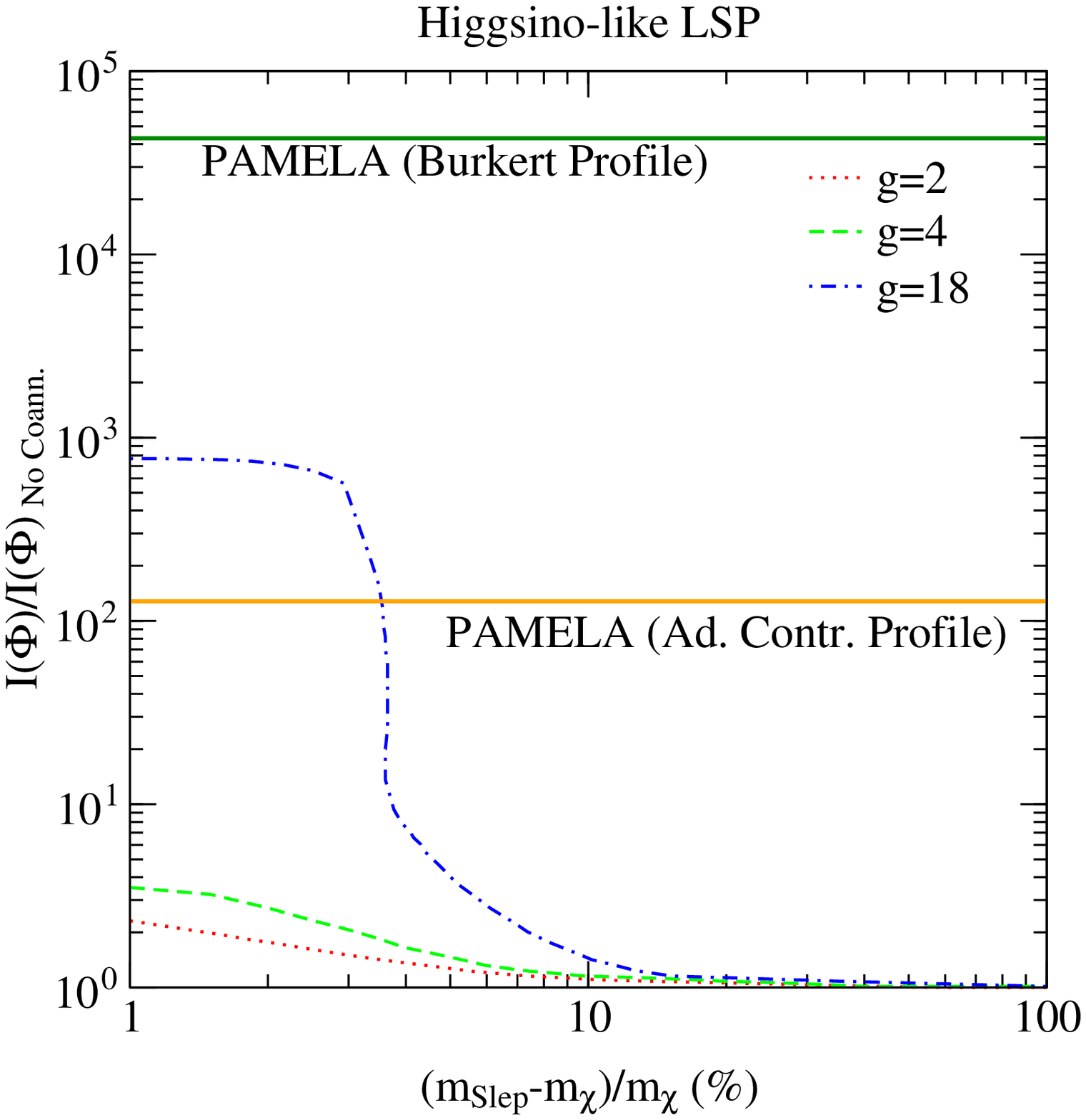,width=7.5cm}\qquad
\epsfig{file=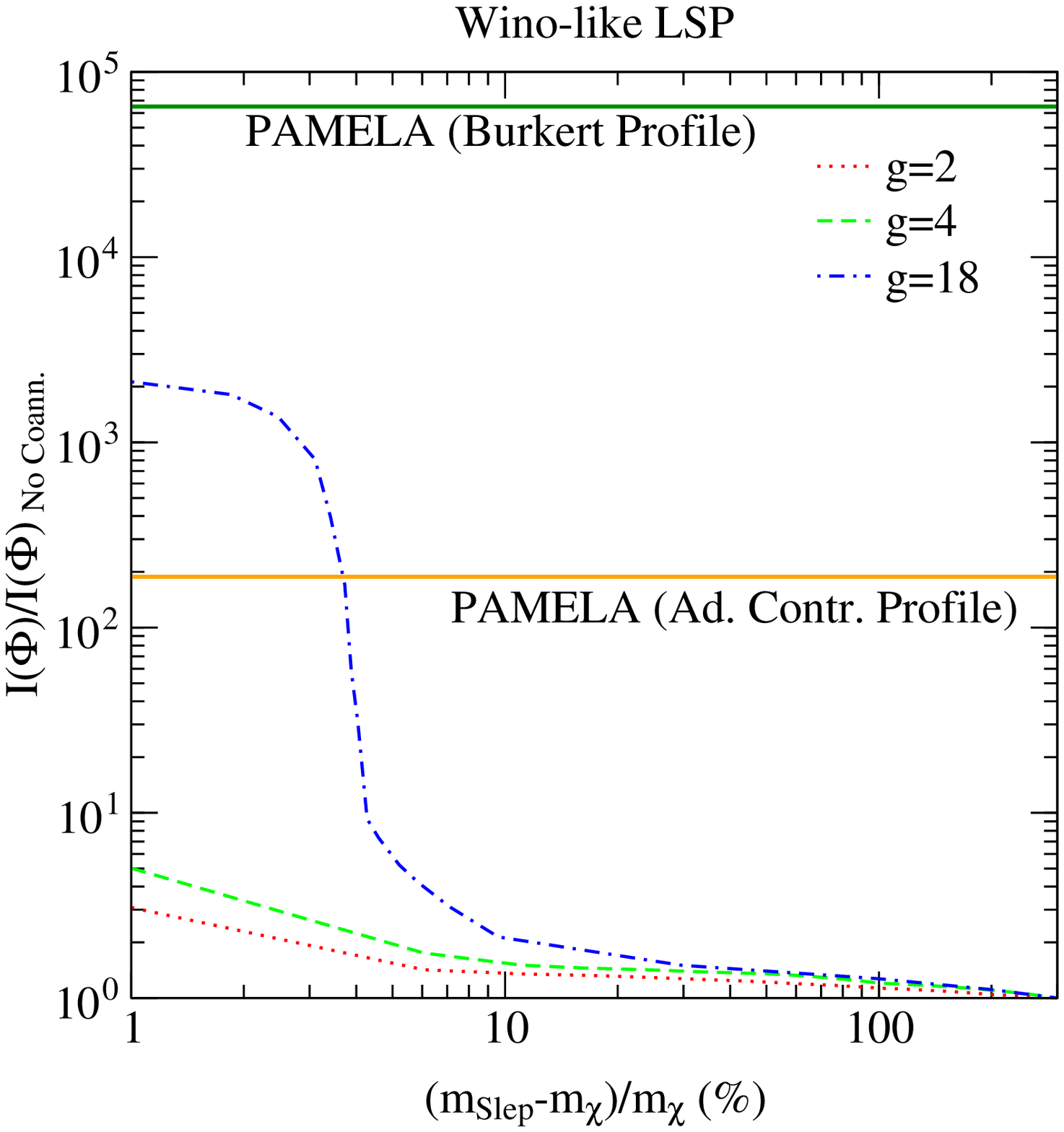,width=7.5cm}}
\caption{\label{fig6}
The relative enhancement (with respect to the asymptotic value with decoupled heavy sleptons)in the quantity $I(\Phi_{\bar p})$, proportional to the expected enhancement of the $\chi^2$ to the antiproton flux with a supersymmetric contribution added on top of the background, as a function of the slepton-LSP percent mass splitting, for  the case of a Higgsino-like neutralino ({left panel}) and  a Wino-like neutralino ({right panel}). The models displayed are those at $\Omega_\chi h^2=0.110$ singled out in Fig.~\protect{\ref{fig3}}, with the same sample choice of parameters and the same line-type and color coding. The horizontal lines indicate the sensitivities of the PAMELA experiment \cite{pamela} after three years of data taking for a cuspy~\cite{n03} and a cored~\cite{burkert} Dark Matter halo profile.}}
The prospects for indirect detection of an exotic signature from galactic Dark Matter annihilations with the recently launched space-based PAMELA experiment~\cite{pamela} are shown in Fig.~\ref{fig6}. We indicate, in the $y$ axis, the enhancement in the antiproton ``{\em Visibility Ratio}'' $I(\Phi)/I(\Phi)_{\rm No\ Coann.}$ where the quantity $I(\Phi)$, first introduced in Ref.~\cite{Profumo:2004ty}, is defined as
\begin{equation}
I(\Phi) \equiv \int_{E_{\rm min}}^{E_{\rm max}} {\rm d}E \,
\frac{\left[\Phi_s(E)\right]^2}{\Phi_b(E)}\;.\label{eq:visibility}
\end{equation}
 $\Phi_s(E)$ and $\Phi_b(E)$ are the signal and background antiproton
fluxes, respectively, at a kinetic antiproton energy $E$, while $I(\Phi)_{\rm No\ Coann.}$ corresponds to the asymptotic case without slepton coannihilations. The quantity $I(\Phi)$ approximates the projected $\chi^2$ of the signal plus background expected with an exotic contribution providing an antiproton flux $\Phi_s(E)$, in the limit of a large number of energy bins \cite{Profumo:2004ty}; $E_{\rm min,\ max}$ indicate the minimal and maximal experimentally accessible antiproton kinetic energies. The treatment of antiproton galactic propagation, diffusion and solar modulation (projected for the actual period of PAMELA data-taking) follows Ref.~\cite{Profumo:2004ty}, where the reader is directed for further details. 
As in the previous plots, we again make use in Fig.~\ref{fig6}, for the $x$ axis, of the  percent mass splitting between the LSP and the coannihilating particles. As shown in Ref.~\cite{Profumo:2004ty} a model gives a statistically significant departure from the background alone, after three years of data-taking and at the 95\% confidence level, if the computed value for $I(\Phi)$ is larger than $3.2\times10^{-8}\,{\rm cm}^{-2}\,{\rm sr}^{-1}\,{\rm s}^{-1}$. We show this sensitivity limit with two horizontal lines, corresponding to a cuspy profile (the adiabatic contraction of the N03 halo model of Ref.~\cite{n03}) and to a cored profile (the Burkert profile \cite{burkert}; for more details on these halo models the reader is directed to Ref.~\cite{Profumo:2004ty,Provenza:2006hr}).

As shown in Fig.~\ref{fig6},  we find a very large enhancement for the case with all sleptons almost degenerate with the lightest neutralino and,  assuming  a cuspy  Dark Matter halo profile~\cite{n03}, PAMELA will be able to statistically disentangle such signal; even with the choice of a cored halo~\cite{burkert} the detection potential of the PAMELA experiment could be sufficient to discriminate an exotic signal, assuming, for instance, a boost factor in the signal flux generated by Dark Matter substructures, or clumps, in the galactic halo, as large as $\approx5$ \cite{substructures}. We moreover wish to point out that we find a similar enhancement in the detection prospects for both positrons and antideuterons, which we do not show here for conciseness, and since it would not add further crucial information to the present discussion. 

Finally, we also computed the expected enhancement in the flux of gamma-rays from neutralino pair annihilations; in this case, we find enhancements very similar to those shown in Fig.~\ref{fig4} for the quantity $\Theta=\langle\sigma v\rangle/m_\chi^2$, integrating the total gamma-ray signal flux in the energy range $E_\gamma>1$ GeV. The question of the actual feasibility of distinguishing a gamma-ray signal originating from neutralino annihilations from the various astrophysical backgrounds relies on several critical assumptions on the Dark Matter distribution and on hypothesis about the background itself, from a given direction in the Sky (see {\em e.g.} the recent discussion given in Ref.~\cite{Zaharijas:2006qb} concerning the case of the Galactic Center). Suffices it to say that if slepton coannihilations are active in the Early Universe at the LSP freeze-out, and if the neutralino is not Bino-like, the shift in the neutralino mass giving the ``right'' thermal relic abundance implies a sizable enhancement (close to what we show in  Fig.~\ref{fig4}) in the expected gamma-ray flux as well.

\section{Conclusions}\label{sec:conclude}

In this paper we studied the effects of slepton coannihilations on the thermal relic abundance of  Higgsino- or Wino-like lightest neutralinos. We pointed out that, unlike the well known case of a Bino-like neutralino, coannihilations with sleptons yield a larger Higgsino and Wino relic abundance. The effect on the relic abundance amounts to an increase of a factor ranging from a few percent up to 5. Requiring that the neutralino relic abundance lies in the range of values preferred for the abundance of Dark Matter entails, in presence of slepton coannihilations, a reduced mass for Winos and Higgsinos, and a larger pair annihilation cross section. Quantitatively, we find that the neutralino mass can be reduced up to a factor between 2 and 3, depending on the particular setup at hand. We showed that smaller values of the neutralino mass and larger pair annihilation cross sections produce potentially very large enhancements in the rates expected in indirect Dark Matter search experiments. In some cases, we showed that the occurrence of slepton coannihilations and the resulting reduction of the neutralino mass needed to produce the right amount of relics is crucial to produce signals that might allow to indirectly probe the occurrence of galactic Dark Matter annihilations.

\acknowledgments

We thank Piero Ullio for valuable discussions and suggestions. The work of S.P. was supported  by the U.S. Department of Energy
grant numbers DE-FG03-92-ER40701 and
FG02-05ER41361, and by NASA grant number NNG05GF69G. The work of A.P. was supported by the Italian INFN under the project ``Fisica Astroparticellare'' and the MIUR PRIN ``Fisica
Astroparticellare''

%

\end{document}